# DISSIPATION IN SINGLE-CRYSTAL 3C-SIC ULTRA-HIGH FREQUENCY NANOMECHANICAL RESONATORS

X. L. Feng,[1] C. A. Zorman,[2] M. Mehregany,[2] and M. L. Roukes[1]

[1]Kavli Nanoscience Institute, California Institute of Technology, 114-36
Pasadena, CA 91125, USA

[2]Electrical Engineering & Computer Science Department, Case Western Reserve University
Cleveland, OH 44106, USA

## ABSTRACT

The energy dissipation $Q^{-1}$ (where $Q$ is the quality factor) and resonance frequency characteristics of single-crystal 3C-SiC ultra-high frequency (UHF) nanomechanical resonators are measured, for a family of UHF resonators with resonance frequencies of 295MHz, 395MHz, 411MHz, 420MHz, 428MHz, and 482MHz. A temperature dependence of dissipation, $Q^{-1} \propto T^{0.3}$ has been identified in these 3C-SiC devices. Possible mechanisms that contribute to dissipation in typical doubly-clamped beam UHF resonators are analyzed. Device size and dimensional effects on the dissipation are also examined. Clamping losses are found to be particularly important in these UHF resonators. The resonance frequency decreases as the temperature is increased, and the average frequency temperature coefficient is about -45ppm/K.

## INTRODUCTION

Nanoelectromechanical systems (NEMS) [1] have attracted considerable research attention and NEMS based resonators offer immense potential for applications such as ultra-sensitive force detection [2], single-molecule mass sensing [3], and mechanical computation [4] and signal processing [5]. In most of these applications, operating at high frequencies provides a key advantage allowing NEMS to surpass conventional solutions. Hence, nanoscale devices and materials with large elastic modulus-to-density ratio ($E/\rho$) are being pursued. Using single-crystal 3C-SiC as a structural material, beam-structured resonators operating in the VHF, UHF, and microwave frequency bands have been demonstrated [6], which are of particular interest for resonant nanomechanical sensing applications. However, one known trade-off is that the quality factor decreases as the resonance frequency increases (i.e., as the devices dimensions are reduced in order to increase the frequency) [6] — thus "Q-engineering" is crucial for retaining high $Q$ while scaling up the frequency. Although the practical performance of these resonators very strongly depends upon quality factors, the mechanisms of dissipation affecting $Q$'s of nanomechanical resonators are still not well understood. It has been qualitatively identified that surface roughness of SiC structural layer can play an important role [6]. With recent advances in ultrahigh quality large SiC crystal growth [7] and surface roughness control [8], SiC promises to be one of the most promising and practical materials for high-performance VHF/UHF/Microwave NEMS. In this work, we report the first systematic measurements and analyses of the dissipation characteristics in several successive generations of single-crystal 3C-SiC UHF NEMS resonators.

## EXPERIMENTAL DETAILS

By following a process specifically suitable for UHF SiC NEMS [6], doubly-clamped beam resonators are nanofabricated from a single-crystal 3C-SiC epitaxial layer grown on a single-crystal silicon substrate by atmospheric pressure chemical vapor deposition (APCVD), supported by newly developed surface roughness control and improvement techniques [8]. Shown in Figure 1 are SEM images of a typical UHF 3C-SiC resonator. The doubly-clamped beam design simplifies understanding of device size and dimensional effects, and also minimizes the influence of complexities in, and variations from, the fabrication processes. Metallization consisting of 10nm Titanium atop a 30nm Aluminum layer is deposited onto the SiC structural material. This enables patterning devices read out by magnetomotive excitation and detection [6] of the beam resonance from the in-plane flexural fundamental mode. Measurements are performed in a strong magnetic field up to $B = 8$T, within a low-temperature cryostat where the device is also secured in high vacuum ($\leq 10^{-7}$ Torr). Typical device dimensions are $w$=120nm, $t$=80nm (excluding metallization layers), and $l$=1.55~2.65µm.

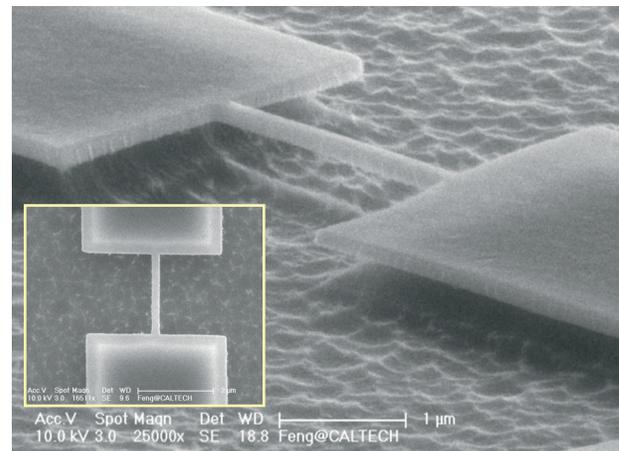

*Figure 1.* Scanning electron micrographs of a typical single-crystal 3C-SiC UHF NEMS resonator. (**main**) Oblique view (scale bar: 1µm) and (**inset**) Top view (scale bar: 2µm).

Network analysis techniques for two-port systems are used to detect the magnetomotively-transduced signals from the devices and to measure the resonance frequencies and quality factors. Resonance signals of these UHF devices are diminutive and are often easily overwhelmed by the embedding parasitic background response arising from the detection system. It has been a challenge to extract large and clean resonance signals out of the electrical background response. We employ a balanced-bridge detection scheme involving the use of pairs of impedance-matched devices [6]. We further incorporate high-resolution bridge-balancing and background-nulling techniques into the circuitry for UHF NEMS resonance readout. These techniques typically allow us to attain signal-to-background ratios (on-resonance to off-resonance ratio) of 5~10dB (whereas previously only 0.1~0.5dB were routinely obtained for UHF NEMS). Quality factors are then reliably extracted from the clean resonance signals of the network analysis measurements for each device under controlled experimental conditions. We note that the same quality factor can also be extracted from time-domain ring-down process of a resonator, albeit somewhat less conveniently than from the frequency-domain power spectrum obtained by network analysis.

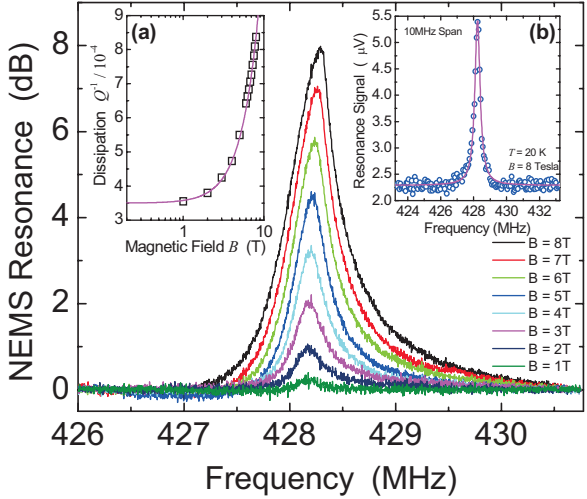

***Figure 2.*** *Resonance signal of a 428MHz NEMS resonator, at various magnetic field conditions, as measured by a microwave network analyzer, utilizing detailed balancing and nulling techniques with a bridge circuitry scheme.* **(inset (a))** *The magnetomotive damping effect.* **(inset (b))** *A typical UHF resonance signal over a 10MHz wide frequency span.*

## RESULTS AND DISCUSSIONS

Figure 2 shows the measured resonance signal of a 428MHz device, with the background response nulled out, as the magnetic field is ramped up from 0T to 8T. The resonance curves clearly indicate an effective $Q$ decrease with increasing magnetic field (as the resonance is getting broadened). The measured dissipation $Q^{-1}$ versus magnetic field is shown in the inset (a) of Figure 2 with the quadratic fit in dashed line. At $B = 8$T, the signal-to-background ratio is 8dB at the resonance peak. The inset (b) of Figure 2 shows the resonance signal referred to the input of the preamplifier, in which both the background signal and the resonance amplitudes are shown in linear scale (in µVolts, 8dB at peak if converted into dB, exactly corresponding to the dB plot in Figure 2; but here in the linear scale plot the absolute level of the flat background, ~2.25µV, *i.e.*, -100dBm, is clearly indicated). The solid line from a Lorentzian fit matches the resonance data over a 10MHz wide span. This readily observed magnetomotive damping effect originates from the fact that the electromagnetomotive force (emf) voltage generated by the vibrating NEMS device in the $B$ field, creates a current as the device is in a closed circuit (with the resistive elements of both the device itself and the measurement system), and in the $B$ field this current induces a force which intends to *oppose* or *damp* the resonating device. This effect can be modelled by a *loaded Q* due to the impedances forming a closed circuit with the emf voltage in the detection system [9],

$$\frac{1}{Q} = \frac{1}{Q_{device}}\left(1 + R_m \frac{\text{Re}(Z_{ext})}{|Z_{ext}|^2}\right), \quad (1)$$

in which $Q_{device}$ is the *unloaded Q* of the device itself; $R_m = Q_{device}\eta B^2 l^2/(2\pi f_0 m)$ is the electromechanical resistance of the device, with $m$ the device mass and mode shape coefficient $\eta=0.83086$ for the fundamental mode; and $Z_{ext}$ (seen by the emf voltage in its closed circuit) is the impedance in series to $R_m$, consisting of the DC resistance of the device, the impedance of the coaxial cable and the input impedance of the preamplifier. Shown in Figure 2 inset (a), the measured $Q$ decrease versus $B$ field, fitting with *Eq.* (1), leads to an estimation of the *unloaded Q* of $Q_{device} \approx 2860$ at ~20K.

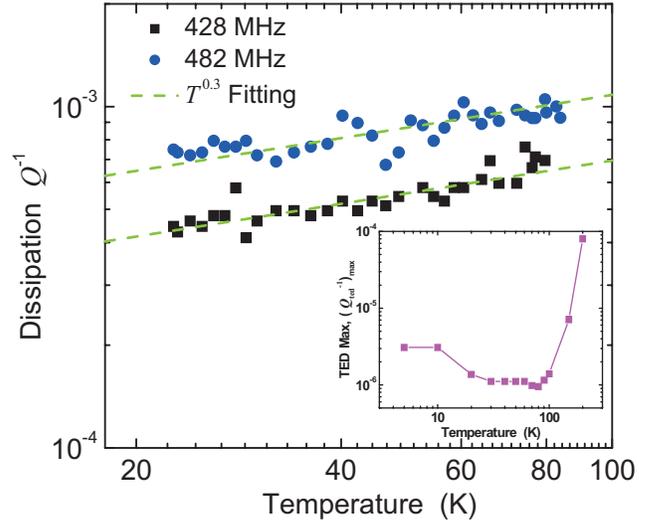

***Figure 3.*** *Measured dissipation as a function of temperature for the selected 428MHz and 482MHz NEMS resonators. The dashed lines show the $Q^{-1} \propto T^{0.3}$ approximation to guide the eyes.* **(inset)** *Theoretical estimation of maximum possible thermoelastic dissipation as a function of temperature.*

To identify the temperature dependence of device dissipation, feedback control of sample temperature is applied and for each measurement, temperature fluctuation is limited to ≤1 mK, in order to minimize the instantaneous resonance frequency variation due to the temperature fluctuation. Figure 3 shows the measured dissipation versus temperature for the 428MHz and 482MHz devices, with all other parameters and settings kept the same and leaving temperature the only variable. RF driving power (-33dBm) with suitable bias field ($B = 6$T) is calibrated and applied to attain large enough signals within the dynamic range in all the temperature dependence measurements. The measured dissipation increases with increasing temperature, with a power-law dependence of roughly $Q^{-1} \propto T^{0.3}$. Note that the data are taken in the range of $T = 20$~100K, which is limited solely by practical limitations of our existing measurement system. Dissipation in resonant nanodevices is complicated and associated with various energy loss mechanisms. Hence, to understand the data requires examinations of all dominant dissipation processes. Assuming that the dissipation from different origins are uncorrelated, the possible important mechanisms that may contribute to the measured dissipation include the 3C-SiC structure layer's intrinsic dissipation $Q_0^{-1}$, magnetomotive damping $Q_{mag}^{-1}$, thermoelastic damping $Q_{ted}^{-1}$, clamping losses $Q_{clamp}^{-1}$, metallization layer dissipation $Q_{metal}^{-1}$, surface losses $Q_{surf}^{-1}$, *etc.*,

$$\frac{1}{Q} = \frac{1}{Q_0} + \frac{1}{Q_{mag}} + \frac{1}{Q_{ted}} + \frac{1}{Q_{clamp}} + \frac{1}{Q_{metal}} + \frac{1}{Q_{surf}} \cdots . \quad (2)$$

Here we neglect the effect of viscous damping in air since all our measurements are performed with devices in high vacuum. The aforementioned, easily observed magnetomotive damping effect can be non-negligible but is readily modelled and identified.

The thermoelastic damping effect arises from the fact that when the beam is deformed and is vibrating, the strain field is coupled to local temperature field and vibratory mechanical energy is dissipated through phonon relaxation processes to the environment. Based on the theory and modelling of thermoelastic damping in doubly-clamped beam resonators [10], we estimate the upper limit of thermoelsatic damping in 3C-SiC devices to be

$$\left(\frac{1}{Q_{ted}}\right)_{max} = \max\left\{\frac{E\alpha^2 T}{\rho C_P}\left[\frac{6}{\xi^2} - \frac{6}{\xi^3}\frac{\sinh\xi + \sin\xi}{\cosh\xi + \cos\xi}\right]\right\}, \quad (3)$$

where $E$, $\alpha$, $\rho$, $C_P$ are the Young's modulus, thermal expansion coefficient, mass density, and heat capacity (per unit mass), respectively. Here $\xi$ is a dimensionless variable representing the relative magnitude of the characteristic size of the device (e.g., device width) versus the characteristic thermal relaxation length (e.g., phonon mean free path). The upper limit of thermoelastic damping, $(1/Q_{ted})_{max}$, independent of the device dimensions, occurs for a system operating at $\xi$=2.225. As plotted in the inset of Figure 3, thermoelastic damping effect in the 3C-SiC NEMS devices is not strong and could be neglected (still more than 2 orders of magnitude smaller than the measured dissipation). We note that this estimation is based on the conventional view in which thermoelastic damping is considered to be a *bulk* effect, and the estimation relies on the available material properties.

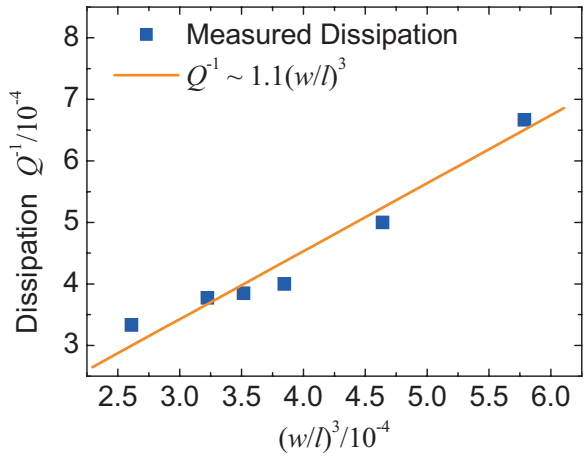

*Figure 4.* Experimental results of dissipation in several generations of UHF NEMS resonators with dimensions and operating frequencies scaled, all measured under the same experimental conditions (the individual device characteristics are listed in Table 1). The solid line is the approximate theoretical fit based on the theory of elastic energy transmission from the vibrating NEMS device to its clamping and supporting pads.

*Table 1.* Specifications of several 3C-SiC UHF NEMS resonators.

| Resonant Frequency (MHz) | Device Dimension | | | Aspect Ratio ($l/w$) | Measured $Q^{-1}$ |
|---|---|---|---|---|---|
| | $l$ (μm) | $w$ (nm) | $t$ (nm) | | |
| 295 | 2.65 | 170 | 80 | 15.65 | 3.33×10$^{-4}$ |
| 395 | 1.75 | 120 | 80 | 14.58 | 3.77×10$^{-4}$ |
| 411 | 1.70 | 120 | 80 | 14.17 | 3.85×10$^{-4}$ |
| 420 | 1.80 | 150 | 100 | 12.00 | 6.67×10$^{-4}$ |
| 428 | 1.65 | 120 | 80 | 13.75 | 4.00×10$^{-4}$ |
| 482 | 1.55 | 120 | 80 | 12.92 | 5.00×10$^{-4}$ |

Scaling resonance frequency up to UHF range brings the doubly-clamped beam length down to only 1~2 μm (e.g., for a SiC layer thickness ~100nm, and a lithography-process-determined width which is typically 100nm~150nm), thus the clamping loss can become important. Theoretical analyses predict that for the in-plane flexural mode of doubly-clamped beams, dissipation into the supports follows $Q_{clamp}^{-1} \approx \beta(w/l)^3$, where the coefficient $\beta$ is not readily modelled [11]. The experimental data from a family of UHF NEMS achieve quite sound agreement with the theoretical prediction, as shown in Figure 4 (with the device characteristics listed in Table 1). The close fit to $Q_{clamp}^{-1} \approx \beta(w/l)^3$ not only indicates that the clamping loss plays an important role that increases with shrinking device dimensions (scaling up frequency); but also suggests that relative role of the clamping loss to the total dissipation, $Q_{clamp}^{-1}/Q^{-1}$, is about the same in these devices. It is thus clear that the offset between the two data traces in Figure 3 is due to the larger clamping loss in the 482MHz device than that in the 428MHz device.

Metallization layers made by evaporation, are not ideal and always possess internal friction, and thus also contribute to the device dissipation. We analyze this by using the general definition $Q^{-1} = \Delta W/(2\pi W)$, where $W$ is the energy stored in the resonator and $\Delta W$ the dissipated energy per cycle, assuming that the energy stored and dissipated can be split into corresponding portions in the structural layer and metallization layers. For our flexural mode doubly-clamped beams, we then obtain

$$\frac{1}{Q} = \sum_{j,k\neq i}^{i=SiC,Al,Ti}\left(1 + \frac{t_j E_j}{t_i E_i} + \frac{t_k E_k}{t_i E_i}\right)^{-1} \cdot \frac{1}{Q_i}, \quad (4)$$

in which $t_i$, $E_i$ are thickness, Young's modulus of the layers (including SiC, Al, Ti), and $Q_i^{-1}$ are the phenomenological dissipation in each layer, respectively. The coefficients for the dissipation in metallization layers are very small, 0.0543 and 0.0293, respectively. With the measured dissipation in deposited submicron Al and Ti films from [12] (each having a plateau in the interested temperature range), the estimated dissipation in metallization layers is 5~6×10$^{-6}$ (as listed in Table 2), still ≤1% of the measured dissipation.

*Table 2.* Estimation of the dissipation contributed by the metallization layers (Al and Ti) with the internal friction in these evaporated metal layers.

| Metal Layer | Thickness (nm) | Young's Modulus (GPa) | $Q_{film}^{-1}$ | Calculated Dissipation |
|---|---|---|---|---|
| Al | 30 | 68 | 1.0×10$^{-4}$ | 5.43×10$^{-6}$ |
| Ti | 10 | 110 | 2.0×10$^{-4}$ | 5.86×10$^{-6}$ |

Surface loss arises from the fact that surface atoms are different than those of the "bulk" and can therefore cause energy dissipation — presumably through forming an additional energy reservoir, or by mediating anharmonic mode coupling, or both. Surface stress, adsorbates and crystal defects on the device surface all may enhance such dissipation. Exact theoretical analyses and models capturing the mesoscopic surface loss mechanisms have yet to be clearly established and are likely often sample-specific and dependent upon *quasi*-random surface conditions. Nonetheless, intuitively one expects larger dissipation (lower $Q$'s) when the device surface-to-volume ratio is increased. Experiments show that for ultra-thin ($t \ll w$) cantilevers, $Q$'s are roughly proportional to the device thickness (in the regime where surface losses are dominant, i.e., when the cantilevers are long enough). This can be qualitatively explained by conventional macroscopic theory based on the concept of a complex modulus ($E=E_1+iE_2$, where $E_2$ is the dissipative part) [13]. For the UHF NEMS devices of very short beams, as shown in Table 1, the surface-to-volume ratio $2(w+t)/(wt)$ does not change much (or almost remains the same) because of fabrication process consistency and the maintenance of relatively constant values for $w$ and $t$. Thus, surface losses in these devices should be approximately the same, and the Q differences measured from these devices are still found to be dominated by clamping losses. To estimate or determine the absolute amount of surface loss, annealing and other surface treatment techniques could, in future, be applied to test how much dissipation might be reduced.

The above analyses show that the observed dissipation temperature dependence $Q^{-1} \sim T^{0.3}$ within the temperature range of these measurements can be ascribed to intrinsic dissipation within the 3C-SiC material itself. Other important mechanisms (*e.g.*, clamping loss) add to this intrinsic dissipation, without markedly changing the temperature dependence. In principle, it may be possible to develop an accurate model describing this temperature dependence with a more detailed, atomistic understanding of the mesoscopic energy dissipation nature within the single-crystal 3C-SiC. The measurements of the temperature-dependent dissipation in 3C-SiC shown herein may provide new insight into this intriguing open question.

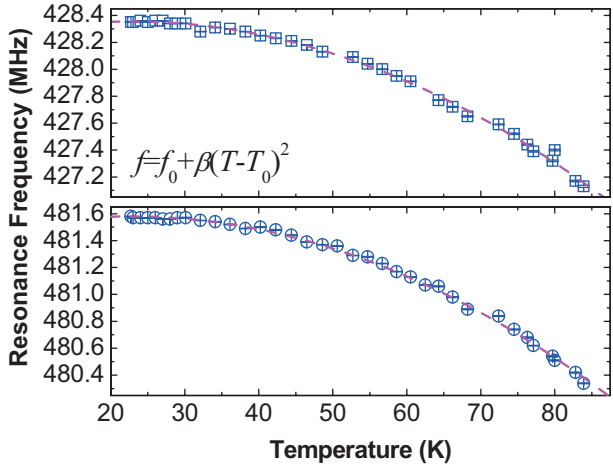

*Figure 5. Measured resonance frequency as a function of temperature for selected 3C-SiC UHF NEMS resonators (at carefully controlled and stabilized temperatures).*

The dependence of resonance frequency upon temperature is measured and is plotted in Figure 5. As shown, the resonance frequency decreases as the temperature is increased. A polynomial fit to the data shows that a quadratic dependence $f = f_0+\beta(T-T_0)^2$ matches the heating-induced ($T \geq T_0$) frequency tuning data quite well, with $\beta \approx -320 Hz/K^2$ for both data traces of Figure 5. We attribute this primarily to the effect of thermal expansion: the frequency changes as both the beam length and its tension are altered with temperature. The results imply that the tensile stress increases monotonically when the devices are cooled down from 85K to 20K. Another observation is that the fractional frequency change does not show dependence upon device size (the two devices shown have different lengths), and both devices exhibit average resonance frequency temperature coefficients of about -45ppm/K in the range $T = 20\sim85K$. This effect, once calibrated over wider temperature range, can be employed to study the basic properties of SiC material, and can be further engineered for sensing applications.

## CONCLUSIONS

We have investigated the dissipation in single-crystal 3C-SiC nanomechanical resonators operating at ultra-high frequencies, to gain understanding and develop engineering solutions that make optimal trade-offs between scaling up resonance frequency and attaining high-$Q$'s for UHF NEMS resonators. It is found that the temperature dependence of the dissipation in the 3C-SiC NEMS resonators studied follows $Q^{-1} \propto T^\alpha$, with $\alpha \approx 0.3$. It is clear that in-depth theoretical models and analyses are needed to reveal the underlying microscopic mechanisms. The magnetomotive damping effect can be appreciable, but it is relatively well understood and (to some extent) controlled. Thermoelastic dissipation is found to be negligible for the devices of this study. The losses from metallization layers contribute ≤1% of the observed total dissipation. However, a major source of the dissipation is clamping losses through the supports for these doubly-clamped beam UHF resonators. The measured data show that the theoretical prediction $Q_{\text{clamp}}^{-1} \propto (w/l)^3$ provides a rough but reasonable model for these clamping losses. Verifying and understanding the dominant clamping losses can lead to new designs and optimization guidelines for UHF NEMS enabling attainment of high $Q$'s. Moreover, because SiC can be deposited both in polycrystalline form as well as in several single-crystal polytypes with excellent properties (including 3C-SiC, 6H-SiC, 4H-SiC and 2H-SiC), it represents a particularly promising material for NEMS applications. Future collective studies of dissipation in SiC NEMS with all these SiC variations would be beneficial.

## ACKNOWLEDGEMENT

This work is supported by the DARPA MTO under grant DABT63-98-1-0012, DARPA/SPAWAR under grant N66001-02-1-8914, and the NSF under grant ECS-0089061.